\documentclass[final]{aipproc} 
\usepackage{graphicx} 
\usepackage{amsmath} 
\usepackage{amssymb} 
\layoutstyle{6x9}
\begin{document} 

\title{Analysis of full-QCD and quenched-QCD lattice propagators}



\author{M.S. Bhagwat}{
address={Center for Nuclear Research, Department of Physics, 
             Kent State University, Kent, Ohio 44242 U.S.A.}
} 
\author{P.C. Tandy}{ 
  address={Center for Nuclear Research, Department of Physics, 
             Kent State University, Kent, Ohio 44242 U.S.A.}
}
%-------------------------------------------------------------------- 

%\date{\today} 

%------------------------------------------------------------------------ 
\begin{abstract} 
Recent lattice-QCD results for the dressed-gluon propagator are used within 
the quark Dyson-Schwinger equation to determine the gluon-quark vertex 
dressing necessary to reproduce the 
lattice-QCD results for the dressed-quark propagator. Both quenched and full 
QCD lattice simulations, for a range of low quark current masses, are 
analyzed.  The chiral extrapolation is made through this continuum DSE form.   
Resulting chiral and physical pion observables are investigated.
\end{abstract} 

%------------------------------------------------------------------------ 
\maketitle 
%------------------------------------------------------------------------ 
\section{Introduction}

Lattice-QCD simulations address hadronic observables via matrix elements of currents
and sources; necessary approximations and truncations introduce systematic  
errors from: a finite volume of discretized space-time, unphysically large current 
quark masses, and the quenched approximation.   While steady progress is being made 
in the reduction of such errors, it is useful to gain insight into dominant field theory
mechanisms by
comparing to hadronic results from covariant modeling of continuum QCD.   The 
Dyson-Schwinger equations (DSEs), the equations of motion of the theory, provide such
an opportunity.  Here systematic errors can arise from truncations that replace
high-order correlations by infrared phenomenology of low-order n-point functions. 
Symmetries can provide significant quality control.   The dressed quark and 
gluon propagators and low-order vertices of the theory are important elements of the
kernels needed for the bound state equations: Bethe-Salpeter equation (BSE) for mesons
and Faddeev equation for baryons.   Prior to the last few years, one had only the 
known ultra-violet behavior of the n-point functions as a guide.  
%-------------------------------------------------------------------------
\begin{figure}[th]
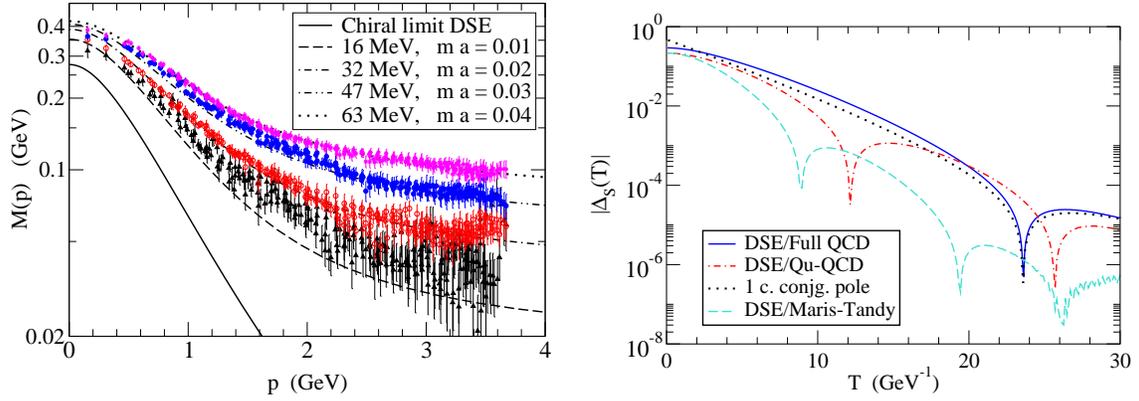

\includegraphics*[width=72mm]{un_massfn_and_fits_proc.eps} 
\hspace{3mm}
\includegraphics*[width=72mm]{FT_MT_qu_full_1cc_proc_Jan4.eps}
\label{Fig:Mfitsandconfine} 
\caption{{\it Left Panel}: Full-QCD lattice results~\protect\cite{Bowman:2005vx}
reproduced from the full-QCD lattice gluon propagator~\protect\cite{Bowman:2004jm}
via the continuum DSE.  Note vertical log scale.   Values of lattice $m$ and $m\, a$
are shown; {\it Right Panel}: Fourier transform of the Dirac
scalar part of the chiral $S(p)$.  Cusps indicate confinement. }
\end{figure}
%------------------------------------------------------------------------

Recently, full-QCD lattice  results for gluon~\cite{Bowman:2004jm} and 
quark~\cite{Bowman:2005vx} propagators have been made available; these are two of the 
three nonperturbative
quantities linked by the quark propagator DSE.   Phenomenological information on the 
third quantity, the dressed quark-gluon vertex $\Gamma^a_\nu(k,p)$, can therefore
be inferred.  We have previously carried out such an analysis of quenched-QCD lattice
propagators~\cite{Bhagwat:2003vw} and the quark-gluon vertex~\cite{Bhagwat:2004kj}; 
the procedures and technical details employed here are described in 
Ref.~\cite{Bhagwat:2003vw}.
Here, and in the lattice simulations, Landau gauge and the Euclidean
metric is used.  Briefly, the approach is the following.  From the renormalized 
quark DSE, we have \mbox{$S^{-1}(p)\!=\!Z_2\, i\gamma \cdot p +$} 
\mbox{$Z_4\, m(\mu) + \Sigma^\prime(p,\Lambda)$} and the regulated self-energy is
\begin{equation}
\Sigma^\prime(p,\Lambda) = Z_1 \! \int^\Lambda_q \!\!g^2\,D_{\mu\nu}(p-q) \, 
\frac{\lambda^a}{2}\gamma_\mu \, S(q) \,\Gamma^a_\nu(p-q,p)~~~.
\label{quarkSelf}
\end{equation} 
Here $\Lambda$ is the regularization mass scale, the $Z_i(\mu^2,\Lambda^2)$ are the
usual constants from renormalization at scale $\mu$.   Two are required so that as
\mbox{$p^2 \to \mu^2$}, we have \mbox{$S^{-1}(p) \to i\gamma \cdot p + m(\mu)$}.
The lattice results for $S(p)$ are given in terms of \mbox{$S(p) = Z(p^2,\mu^2)/$}
\mbox{$(i\gamma \cdot p + M(p^2))$}.   We compare by solving \eqref{quarkSelf} with
the kernel factors \mbox{$ Z_1 \,g^2\,D_{\mu\nu}(k)\,\Gamma^a_\nu(k,p)$} replaced by
\mbox{$D_{\mu\nu}^{\rm lat}(k)\,\frac{\lambda^a}{2}\, \gamma_\nu \, \Gamma_1(k^2)$}.
Here $D_{\mu\nu}^{\rm lat}(k)$ is a fit to the lattice gluon propagator and the 
quantity $\Gamma_1(k^2)$ is a phenomenological vertex amplitude determined so that 
the DSE solution for $S(p)$ fits the lattice data.   In general 
the transverse vertex has eight amplitudes and a dependence upon two momenta; 
however the data being fitted here do not warrant more.   We note that one may 
identify the kernel as $4 \pi \alpha_{\rm eff}(k^2)\, D_{\mu\nu}^0 (k)$ where
$D_{\mu\nu}^0$ is the $0^{\rm th}$ order gluon propagator.   We ensure that the
leading log behavior of all quantities conform to the 1-loop renormalization group 
behavior of QCD.       

\section{Results}

%-----------------------------------------------------------------------------------------
\begin{table}[ht] 
\caption{Chiral condensate and pion decay constant (chiral $f_\pi^0$, physical $f_\pi$)
from the lattice-guided DSE kernel.}
\begin{tabular}{lccc} \hline
                                            &  Expt       & Full-QCD   & Qu-QCD \\ \hline 
$\langle \bar{\!\!\!q\,} q\rangle_{~~\mu=1~{\rm GeV}}$ & 
                               -(0.24~GeV)$^3$ & -(0.23~GeV)$^3$ & -(0.19~GeV)$^3$ \\ \hline 
$f_{\pi}^0$                    & 0.090~GeV    &  0.072~GeV     &  0.063~GeV     \\ \hline 
$f_{\pi}$                      & 0.092~GeV    &  0.075~GeV     &  0.066~GeV     \\ \hline 
\end{tabular}
\label{tab:obsv}
\end{table}
%------------------------------------------------------------------------------------------
Our fit to the full-QCD $D_{\mu\nu}^{\rm lat}(k)$ uses the ``model A'' form previously 
employed for quenched data~\cite{Leinweber:1998uu} but now with \mbox{$N_f =3$} and
the new parameter values: $A = 3.25$, $\Lambda_{\rm g} = 0.54$, $\alpha = 1.15$ and 
$Z_{\rm g} = 1.22$.   Our fit gave priority to $M(p)$ for the four available values of lattice
$m\,a$ shown, along with the results, in the left panel of 
Fig.~\ref{Fig:Mfitsandconfine}.   The parameterized form used for $\Gamma_1(k^2)$ is the same
as we have previously used~\cite{Bhagwat:2003vw} for the quenched lattice case.   
The (dimensionless) parameters of $\Gamma_1(k^2)$ found here for the full-QCD case are: 
$a_1 = 4.5, a_2 = 2.1, a_3 = 18.1, b = 0.31$.   The result of the  chiral 
extrapolation provided by the DSE kernel is also shown.
The lattice current masses are equally spaced and, in the region $p \gtrsim 3 GeV$, 
the lattice results are only approximately so; the $m\,a =0.01$ 
case deviates most from the pattern.   The DSE fit has been made with the constraint
that the $M(p,m)$ approach the correct ratio. 
  
If a propagator of a field theory in Euclidean metric violates the Osterwalder-Schrader 
axiom of relection positivity~\cite{Osterwalder:1974tc}, then this is a sufficient 
condition for 
confinement of the corresponding excitation~\cite{Krein:1992sf}.  In the right panel of 
Fig.~\ref{Fig:Mfitsandconfine} we display the magnitude of the Fourier transform of 
$\sigma_s(p_4, \vec{p}=0)$, the Dirac scalar amplitude of the chiral limit quark propagator.  
For a free particle with mass $m$, \mbox{$\Delta_S(T) \propto {\rm exp}(-mT)$}.   
The cusps indicate changes of sign and thus confinement in both quenched and full-QCD. 
The dotted line corresponds to a propagator that has a single pair of complex conjugate poles at 
\mbox{$p^2 = - 0.309^2 \pm i 0.192^2~{\rm GeV}^2 $}.   The dashed curve corresponds to
the ladder-rainbow model that describes a large variety of light quark 
observables~\cite{Maris:1999nt}. 

The form of the deduced DSE kernel allows a chiral symmetry-preserving Bethe-Salpeter
kernel to be obtained as in the ladder-rainbow case~\cite{Bender:1996bb}.   This
produces the chiral physics observables shown in Table~\ref{tab:obsv}.
Full-QCD evidently produces an acceptable condensate, while the quenched 
approximation underestimates by a factor of two~\cite{Bhagwat:2003vw}.
The values of $f_\pi$, both chiral and physical, are marginally improved by full-QCD
but they remain about 15\% too low.

\begin{theacknowledgments}
Conversations with Craig Roberts and Pieter Maris have been valuable.  This work has 
been partially supported by NSF grant no. PHY-0301190.
\end{theacknowledgments}

% Mandar \bibliography{PANIC05proceedings}

%----bibtex-------------------------------------------------
\bibliographystyle{aipproc}   % if natbib is available

%% use your own bibtex database(s) here
\bibliography{refsPM,refsPCT,refsCDR,refs}
%-----------------------------------------------------------

\end{document}